\documentclass{iaus}
\usepackage{graphicx}
\newcommand{\apj}{\textit{ApJ}}
\newcommand{\aj}{\textit{AJ}}

\newcommand{\apjl}{\textit{ApJ Lett.}}
\newcommand{\apjs}{\textit{ApJ Suppl.}}

\newcommand{\mnras}{\textit{MNRAS}}

\title[] 
{Voids in the Local Volume: a limit on
appearance of a galaxy in a DM halo }

\author[Tikhonov \& Klypin]   
{Anton V. Tikhonov$^1$, Anatoly A. Klypin$^2$%
  \thanks{Present address: Konstantinova street,  4, app. 40, 188308, Gatchina,
Leningradsky region, Russia.},
  }

\affiliation{$^1$Chair of Astrophysics, Department of Mathematics
and Mechanics, St.Petersburg State University, Universitetsky
prospect, 28, Saint-Petersburg, Petrodvoretz,
198504 Russian Federation \break email: ti@hotbox.ru, avt@gtn.ru\\[\affilskip]
$^2$ The Astronomy Department, New Mexico State University,
\break Las Cruces, New Mexico 88003-8001,
USA \break email: aklypin@nmsu.edu}

\pubyear{2007}
\volume{244}  
\pagerange{000--000}
\date{July 23, 2007}
 \jname{Proceedings Dark Galaxies and Lost
Barions IAU Symposium} \editors{A.C. Editor, B.D. Editor \& C.E.
Editor, eds.}
\begin{document}

\maketitle

\begin{abstract}

Current explanation of the overabundance of dark matter subhalos in
the Local Group (LG) indicates that there maybe a limit on mass of a
halo, which can host a galaxy. This idea can be tested using voids in
the distribution of galaxies: at some level small voids should not
contain any (even dwarf) galaxies.  We use observational samples
complete to $M_B=-12$ with distances less than 8~Mpc to construct
the void function (VF): the
distribution of sizes of voids empty of any galaxies. There are $\sim
30$ voids with sizes ranging from 1 to 5~Mpc.  We then study the
distribution of dark matter halos in very high resolution simulations
of the LCDM model. The theoretical VF matches the observations
remarkably well only if we use halos with circular velocities larger
than $45\pm 10$~km/s.  This agrees with the Local Group
predictions. There are smaller halos in the voids, but they should not
produce any luminous matter.  Small voids look quite similar to their
giant cousins: the density has a minimum at the center of a void and
it increases as we get closer to the border.  Small nonluminous halos
inside the void form a web of tiny filaments. Thus, both the Local
Group data and the nearby voids indicate that isolated halos below
$45\pm 10$~km/s must not host galaxies and that small (few Mpc) voids
are truly dark.

\keywords{galaxies: structure, statistics, halos;
 cosmology: large-scale structure of universe, dark matter}
\end{abstract}

\firstsection 
\section{Introduction}

The observational discovery of giant voids (\cite{Gregory1978};
\cite{Joeveer1978}; \cite{Kirshner1981})
was soon followed by the theoretical understanding that voids
constitute a natural outcome of structure formation via gravitational
instability (\cite{Peebles1982}; \cite{Hoffman1982}). Emptiness of
voids -- the number of small galaxies in the voids -- is an
interesting question for both the observations and the theory to
tackle: do we have a problem (\cite{Peebles2001})? Cosmological
simulations predict (e.g., \cite{Gottloeber2003})  that many small
DM halos should reside in voids. There seems to be no disagreement
between the LCDM theory and the observations (\cite{Patiri2006})
 regarding the giant voids defined by $M_*$ galaxies or by
 $10^{12}M_{\odot}$ halos. The situation is less clear on smaller
 scales. In the region of $\sim 10$~Mpc around the Milky Way, where
 observations go to remarkably low luminocities, small voids look very
 empty: dwarf galaxies do not show a tendency to fill the voids and
 voids are still relatively large.  The theory predicts that many
 dwarf dark matter halos should be in the voids, which puts it in the
 collision course with observations.  Yet, below some mass the halos
 are expected to stop producing galaxies inside them. There are
 different arguments for that: stellar feedback (\cite{Dekel1986}) or
 photoionozation may play significant role in quenching star formation
 in too small halos. Still, it is difficult to get a definite answer
 because the physics of dwarfs at high redshifts is quite complicated.

Satellites of the Local Group give a more definite answer. Current
explanation of the overabundance of the dark matter subhalos
(\cite{Kravtsov2004}) assumes that dwarf halos above $V_c \approx
50$~km/s were forming stars before they fall into the Milky Way or
M31. Once they fall in, they get severely striped and may
substantially reduce their circular velocity producing galaxies such
as Draco or Fornax with the rms line-of-sight velocities only few
km/s.  The largest subhalos retain their gas and continue form stars,
while smaller ones may lose the gas and become dwarf
spheroidals. Halos below the limit never had substantial star
formation. They are truly dark.  This scenario implies that
$V_c\approx 50$~km/s is the limit for star formation in halos. If this
picture is correct, it can be tested with small-size voids: they must
be empty of any galaxies
 and are filled with gas and dark matter halos.

\section{DATA: Local Volume}\label{sec:data}

The first catalog of galaxies within 10~Mpc was compiled by
\cite{Tammann1979} and contained 179 galaxies.  Tully (1988) noted
that the Local Supercluster contains a number of filaments and
that those outline the so-called Local Void, which begins just
outside the Local Group and extends in the direction of the North
Pole of the LSC. The Local Void looks practically free from
galaxies. Over the past few years special searches for new nearby
dwarf galaxies have been undertaken using numerous observational
data: the optical sky survey POSS-II/ESO/SERC, HI, and on infrared
surveys of the zone of avoidance, HIPASS and HIJASS. At present,
the sample of galaxies with distances less than 10~Mpc lists about
500 galaxies. For half of them the distances have been measured to
an accuracy as high as 8-10\% (\cite{Karachentsev2004}). Over the
last 5~years snapshot surveys with Hubble Space Telescope (HST)
have provided us with the TRGB distances for many nearby galaxies.
The absence of the ``finger of God'' effect in the Local Volume
simplifies the analysis of the shape and orientation of nearby
voids. Observations of the Local Volume have detected dwarf
systems down to extremely low luminosity. This gives us unique
possibility to detect voids which may be empty of any galaxies.
\cite{Tikhonov2006} analyzed nearby voids. Here we continue the
analyzis using an updated list of galaxies (Karachentsev, private
communication).  The volume limited sample is complete for
galaxies with abs. magnitudes $M_B=-12$ within 8~Mpc radius.

\section{Simulations}
We use N-body simulations done with the Adaptive Refinement Tree code
(\cite{Kravtsov1997}).  The simulations are for spatially flat
cosmological LCDM model with following parameters:
$\Omega_0 = 0.7,  \Omega_{\Lambda} = 0.3; \sigma_8 = 0.9;
H_0 = 70$ km/s/Mpc.
As a measure of how large is a halo we typically use the maxumum
circular velocity $V_c = (GM/R)^{1/2}$. This quantity  is
easier to relate to observatons as compared with the virial mass.
For reference, halos with $V_c =50$~km/s have virial mass about $10^{10}M_{\odot}$
and halos with  $V_c =20$~km/s have virial mass about $10^{9}M_{\odot}$.
We use two simulations: (1) Box 80Mpc/h (Box80); mass per particle
$3\times 10^8h^{-1}M_{\odot}$; simulations cover the whole volume and (2) Box 80Mpc/h (Box80S);
 spherical region of 10~Mpc inside 80Mpc/h box resolved with
 $5\times 10^6h^{-1}M_{\odot}$ particles. We use halos resolved at least with
 20~particles.

\section{Detecting Voids}\label{sec:voids}
In order to detect voids, we place a 3d mesh on the observational or simulation volume.
We then find initial centers of voids as the mesh centers having the largest distances to
nearest objects. In the next iteration, an initial spherical void may be increased by
adding additional off-center empty spheres with smaller radius. The radius of the spheres
 is limited to be larger then 0.9 of the initial sphere and their centers must stay
inside the volume of the first sphere. The process is repeated few times. It produces
voids which are slightly aspherical, but voids never become more flattened than 1:2 axial ratio.
Artificial objects are palced on the boundaries of the sample to prevent voids getting
 out of the boundaries of the sample.
We define the cumulative void function (CVF) as the fraction of the total volume occupied
 by voids with effective radius larger than
$R_{\rm eff} = (3 V_{\rm void}/4\pi)^{-1/3}$.

\section{Cumulative Void Function of the  Local Volume}\label{sec:cvf}

In order to construct CVF of the Local Volume we use two samples: (1)
Galaxies brighter than $M_B=-12$ inside sphere of radius 8~Mpc and (2)
all galaxies inside 7.5~Mpc. Results are present in the right panel of
Figure 1. There are about 30~voids in the observational sample. We
limit the radius of voids to be more than 1~Mpc. The two subsamples
indicate some degree of stability: inclusion of few low-luminocity
galaxies does not change the void function.

We use the Box80 simulation (full volume) to  constract a sample of 40  ``Local Volumes''.
The selecton criteria are:  1) no
halos with $M > 10^{14}M_{\odot}$ inside a  8~Mpc sphere (thus, no clusters
in a sample); 2) the sphere must be centered on a halo with $150 < V_c  < 200$
km/s (Milky Way analog). Because of the mass resolution, the halo catalogs
 are complete down to halos with circular velocity $V_c=40$~km/s.
The second simulation (Box80S) provides one sample and it is complete down to 20~km/s.

The left panel in the Figure 1 shows CVF for different samples of
halos and the observed CVF.  Results indicate that voids in the
distribution of halos with $V_c > 45$~km/s give the best fit to the observed CVF.
The theoretical  CVF goes above the observational
data if we use circular velocities larger than 60~km/s. If we use
significantly lower limits, than the theory predicts too few large
voids. The theoretical results match the observations
if we use $V_{circ} = 45 \pm 10$~km/s. In this case, the match is
 remarkably good: the whole spectrum of voids is reproduced by the theory.

\begin{figure}
 \includegraphics[scale=0.65]{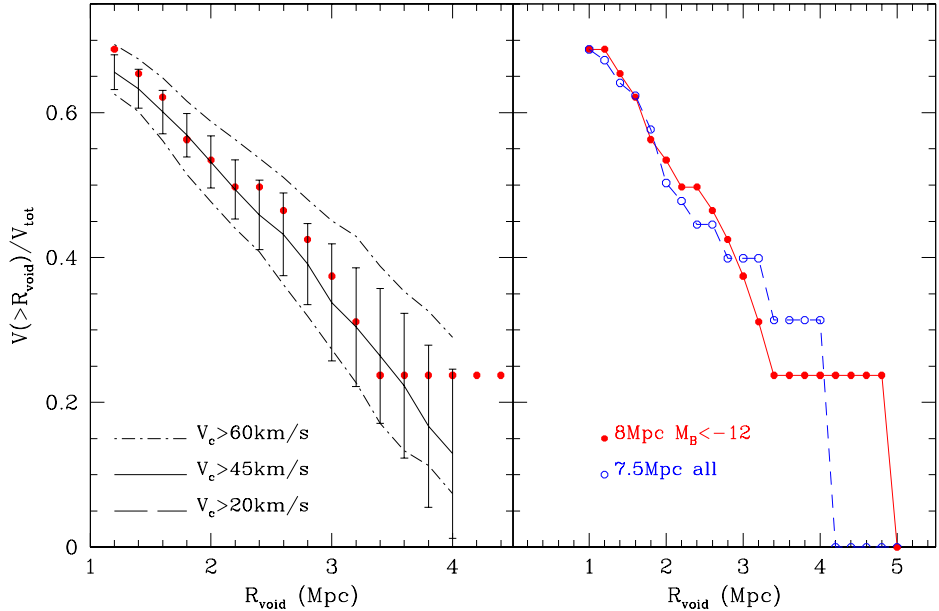}
  \caption{Right panel:
the void function for two observational samples. The full curve and
filled circles are for a complete volume limited sample with $M_B<-12$
and $R<8$~Mpc.  The open circles are for all oserved galaxies inside
7.5~Mpc. Comparison of the samples shows reasonable stability of the
void function. Left panel: Observational data (the complete sample)
are compared with the distribution of voids in samples of halos with
different limits on halo circular velocity. CVF for $V_c=45$~km/s
provides a remarkably good fit to observations. Note that the LCDM
model predicts very large empty regions.
}\label{fig:rms}
\end{figure}
According to LCDM simulations totally empty front part of the Local
Void is probable. In a sample  of ten 8~Mpc "Local Volumes" a half of
cases have voids comparable to the largest voids in LV if we consider entire
LV sample.

\section{Interiors of voids in LCDM}\label{sec:torustrans}
Density  of matter inside voids increases toward borders of voids.
This indicates that voids are
physical; not statistical flukes because the same behaviour was found for
giant voids in simulations of \cite{Gottloeber2003}. To large degree
the small and giant voids are similar. For example, there are very small
 filaments made of tiny halos inside small voids.
Small voids can contain very few small halos.
When studiing voids defined by $V_c> 45$~km/s, we often find relatively
large voids (few Mpc radius) totally empty of halos with $V_c > 20$~km/s.

\section{Luminosity funcion of LV}\label{sec:lf}

The luminocity function of galaxies in the Local Volume, complete sample {$M_B < -12$}, is peaked on $M_B
\sim -14$. The peak is mostly produced by isolated galaxies. Though
it's statistical significance is still questionable, galaxies of this luminosity
can be associated with DM halos with $V_c \sim 35-40$~km/s.
Taking into accout our main result - limit of appearance on
$V_c = 45 \pm 10$~km/s the bump in LF may be real. Better data
are needed to make more definite conclusions.
In particular, measurements of the rotation velocity function of LV galaxies is
needed to find if there is something special about galaxies, which are close to
the limit $V_c=30-50$~km/s. They are isolated  dwarf irregulars, which have
large gas mass. Is it enough to slightly boost their luminocity and produce a maxium
in the luminocity function before it dramatically decreases for smaller halos?

\section{Conclusions}\label{sec:concl}

We find that
\begin{itemize}
\item
The LCDM model is consistent with the cumulative volume functions of voids in
the distribution of galaxies for a large
luminosity range. According to LCDM, large empty voids in Local Volume such as
the Local Void are probable.

\item
There are significant (up to few Mpc)  holes in
the distribution of halos predicted by LCDM that are free from haloes
with $V_c > 20$~km/s: any  haloes of astronomical interest.

\item
Voids in the distribution of haloes with $V_c > 45 \pm 10$~km/s
reproduce the Cumulative Void Function of Local Volume galaxy sample.
We can treat this value as a limit of appearance of a galaxy in
a DM halo.

\item
$M_B \sim -14$ isolated galaxies (probably having $V_c \sim
35-45$~km/s) may be on the limit of appearance. We speculate that the peack in
Luminosity Function at this magnitude may be real. More data is needed to test this conclusion.

\item
Dark halos are probably located  close to borders of voids.

\end{itemize}

\begin{acknowledgments}
We thank I.D.~Karachentsev for providing us an updated list of his
Catalog of Neighboring galaxies. A.~Klypin acknowledges support of
NSF grants to NMSU. Computer simulations used in this research
were conducted on the Columbia supercomputer at the NASA Advanced
Supercomputing Division and at the Leibniz-Rechenzentrum (LRZ),
Munchen, Germany. A.~Tikhonov acknowledges support of grant no.
MK-6899.2006.2 from the President of Russia.

\end{acknowledgments}

\begin{discussion}

\end{discussion}


\begin{thebibliography}{}

\bibitem[Dekel \& Silk(1986)]{Dekel1986} Dekel, A., \& Silk, J.\
1986, \apj, 303, 39

\bibitem[Gottl{\"o}ber et al.(2003)]{Gottloeber2003} Gottl{\"o}ber,
S., {\L}okas, E.~L., Klypin, A., \& Hoffman, Y.\ 2003, \mnras, 344, 715

\bibitem[Gregory \& Thompson(1978)]{Gregory1978} Gregory, S.~A., \&
Thompson, L.~A.\ 1978, \apj, 222, 784

\bibitem[Hoffman \& Shaham(1982)]{Hoffman1982} Hoffman, Y., \&
Shaham, J.\ 1982, \apjl, 262, L23
\bibitem[Peebles(1982)]{Peebles1982} Peebles, P.~J.~E.\ 1982, \apj,
257, 438

\bibitem[Joeveer \& Einasto(1978)]{Joeveer1978} Joeveer, M., \&
Einasto, J.\ 1978, Large Scale Structures in the Universe, 79, 241
\bibitem[Kirshner et al.(1981)]{Kirshner1981} Kirshner, R.~P.,
Oemler, A., Jr., Schechter, P.~L., \& Shectman, S.~A.\ 1981, \apjl, 248,
L57

\bibitem[Karachentsev et al.(2004)]{Karachentsev2004} Karachentsev,
I.~D., Karachentseva, V.~E., Huchtmeier, W.~K., \& Makarov, D.~I.\ 2004,
\aj, 127, 2031

\bibitem[Kraan-Korteweg \& Tammann(1979)]{Tammann1979}
Kraan-Korteweg, R.~C., \& Tammann, G.~A.\ 1979, Astronomische Nachrichten,
300, 181

\bibitem[Kravtsov et al.(1997)]{Kravtsov1997} Kravtsov, A.~V.,
Klypin, A.~A., \& Khokhlov, A.~M.\ 1997, \apjs, 111, 73


\bibitem[Kravtsov et al.(2004)]{Kravtsov2004} Kravtsov, A.~V.,
Gnedin, O.~Y., \& Klypin, A.~A.\ 2004, \apj, 609, 482

\bibitem[Patiri et al.(2006)]{Patiri2006} Patiri, S.~G., Prada,
F., Holtzman, J., Klypin, A., \& Betancort-Rijo, J.\ 2006, \mnras, 372,
1710

\bibitem[Peebles(2001)]{Peebles2001} Peebles, P.~J.~E.\ 2001, \apj,
557, 495

\bibitem[Tikhonov \& Karachentsev(2006)]{Tikhonov2006} Tikhonov,
A.~V., \& Karachentsev, I.~D.\ 2006, \apj, 653, 969

\bibitem[Tully \& Fisher(1987)]{Tully1987} Tully, R.~B., \&
Fisher, J.~R.\ 1987, Atlas of Nearby Galaxies, Cambridge University Press.



\end{thebibliography}
\end{document}